# A white dwarf cooling age of 8 Gyr for NGC 6791 from physical separation processes

Enrique García-Berro[1,2], Santiago Torres[1,2], Leandro G. Althaus[1,3,4], Isabel Renedo[1,2], Pablo Lorén-Aguilar[1,2], Alejandro H. Córsico[3,4], René D. Rohrmann[5], Maurizio Salaris[6], Jordi Isern[2,7]

[1]Departament de Física Aplicada, Universitat Politècnica de Catalunya, c/Esteve Terrades 5, 08860 Castelldefels, Spain

[2]Institut d'Estudis Espacials de Catalunya, Ed. Nexus-201, c/Gran Capità 2-4, 08034 Barcelona, Spain

[3]Facultad de Ciencias Astronómicas y Geofísicas, Universidad Nacional de La Plata, Paseo del Bosque s/n, 1900 La Plata, Argentina

[4]Instituto de Astrofísica de La Plata (CCT La Plata), CONICET

[5]Instituto de Ciencias Astronómicas, de la Tierra y del Espacio, CONICET, Av. de España 1512 (Sur) CC 49, 5400 San Juan, Argentina

[6]Astrophysics Research Institute, Liverpool John Moores University, 12 Quays House, Birkenhead, CH41 1LD, United Kingdom

[7]Institut de Ciències de l'Espai (CSIC), Facultat de Ciències, Campus UAB, Torre C5-parell, 2ª planta, 08193 Bellaterra, Spain

**NGC 6791 is a well studied open cluster[1] that it is so close to us that can be imaged down to very faint luminosities[2]. The main sequence turn-off age (~8 Gyr) and the**

**age derived from the termination of the white dwarf cooling sequence (~6 Gyr) are significantly different. One possible explanation is that as white dwarfs cool, one of the ashes of helium burning, $^{22}$Ne, sinks in the deep interior of these stars[3,4,5]. At lower temperatures, white dwarfs are expected to crystallise and phase separation of the main constituents of the core of a typical white dwarf, $^{12}$C and $^{16}$O, is expected to occur[6,7]. This sequence of events is expected to introduce significant delays in the cooling times[8,9], but has not hitherto been proven. Here we report that, as theoretically anticipated[5,6], physical separation processes occur in the cores of white dwarfs, solving the age discrepancy for NGC 6791.**

White dwarf stars are the most common end-point of stellar evolution. Because they are very old objects, they convey important information about the properties of all galactic populations, including globular and open clusters. This is particularly true for NGC 6791, a metal-rich ([Fe/H]~+0.4), well populated (~3000 stars) and very old (~8 Gyr) open cluster that has been imaged down to luminosities below those of the faintest white dwarfs[1,2], thus providing us with a reliable white dwarf luminosity function[2]. The white dwarf luminosity function of NGC 6791 presents two prominent peaks (a rather peculiar feature). The first of these peaks has been interpreted as either the result of a population of unresolved binaries[10], or a population of single helium white dwarfs[11]. The second peak and the subsequent drop-off in the white dwarf luminosity function are a consequence of the finite age of the cluster. The age obtained using the white dwarf luminosity function of NGC 6791 is in conflict with the age of the cluster derived from its main sequence stars. This discrepancy cannot be attributed to a poor determination of the main sequence turn-off age, since for this cluster we have a reliable determination of its age, 8.0±0.4 Gyr. The age uncertainty mainly arises from the uncertainty in the metallicity determination and is probably an overestimate[2]. Several explanations for solving this discrepancy have been proposed[1,2]. Amongst them a different distribution of carbon and oxygen in the cores of white dwarfs, $^{22}$Ne sedimentation and carbon-oxygen phase separation upon crystallisation have been proposed. Of these, the most viable and promising explanation is, precisely, a combination of the last two ones[2,5,9], since the high metallicity of this cluster makes these effects much more important. However, no white dwarf cooling sequences incorporating both $^{22}$Ne sedimentation and carbon-oxygen phase separation upon crystallisation were available until now, thus hampering a confirmation of this hypothesis. To this end, we have followed the entire evolution of 0.5249, 0.5701, 0.593, 0.6096, 0.6323, 0.6598 and 0.7051 $M_\odot$ white dwarf sequences which include both physical processes.

Our sequences start from stellar models on the zero-age main sequence with masses between 1 and 3 $M_\odot$. These sequences were followed through the thermally pulsing and mass-loss phases on the asymptotic giant branch to the white dwarf stage. Evolutionary calculations were done using a state-of-the-art stellar evolutionary code[12]. Issues such as the simultaneous treatment of non-instantaneous mixing and burning of elements, and the modelling of extra-mixing episodes during the core nuclear burning, of relevance for the carbon-oxygen stratification of the resulting white dwarf core, have been considered with a high degree of detail. Particularly relevant for the present study is the treatment of the release of gravitational energy resulting from $^{22}$Ne sedimentation in the liquid phase and from the phase separation of carbon and oxygen upon crystallisation[13,14]. At the evolutionary stages where the fainter peak of the white dwarf luminosity function of NGC 6791 is observed, these effects markedly slow down the cooling process of white dwarfs. The inclusion of these two energy sources is done self-consistently, and locally coupled to the full set of equations of stellar evolution. The energy contribution of $^{22}$Ne sedimentation was computed assuming that the liquid behaves as a single background one-component plasma characterized by the number average of the real carbon and oxygen one[8], plus traces of $^{22}$Ne. In this way we assess the diffusively evolving $^{22}$Ne profile in a simple and realistic manner. The diffusion coefficient of $^{22}$Ne was the theoretical one[5]. The energy contribution arising from core chemical redistribution upon crystallisation was computed keeping constant the abundance of $^{22}$Ne, in accordance with theoretical calculations[15]. We adopted a carbon-oxygen phase diagram of the spindle form[16]. Finally, the constitutive physics of our model comprise updated radiative and conductive opacities, neutrino emission rates, a detailed equation of state for both the liquid and solid phases and realistic boundary conditions for cool white dwarfs, as given by non-grey model atmospheres. Calculations were conducted down to very low surface luminosities, well beyond the luminosity corresponding to the fainter peak of the white dwarf luminosity function of NGC 6791.

We simulated the white dwarf luminosity function of NGC 6791 using a Monte Carlo technique[17,18,19]. Synthetic main sequence stars were randomly drawn according to a standard initial mass function with exponent −2.35, and a burst of star formation which lasted for 1 Gyr, occurring 8 Gyr ago. We accounted for unresolved detached binary white dwarfs by considering a total binary fraction equal to 54%, with the same distribution of secondary masses, as previous studies did[10]. This overall binary fraction leads to a 36% of white dwarf binary systems on the cooling sequence. The main sequence lifetimes were obtained from up-to-date evolutionary calculations[20] for the metallicity of NGC 6791, and we used an initial-to-final mass relationship appropriate for metal-rich stars[21]. Given the age of the cluster, the time at which each star was born and its main sequence lifetime, we were able to determine which stars entered the white dwarf cooling track, which were their cooling times and we interpolated the luminosity and colours in the theoretical cooling sequences. If the star belongs to an unresolved binary system we did the same calculation for the secondary and we added the fluxes. We also considered photometric errors according to Gaussian distributions. The standard photometric error was assumed to increase linearly with the magnitude[1,2]. Finally, we took into account the distance modulus of NGC 6791, $(m-M)_{F606W}=13.44$, and its colour excess, $E(F606W-F814W)=0.14$, as derived from the most recent observations[1,2,22]. Following all these steps we were able to produce a synthetic colour-magnitude diagram.

An example of a typical Monte Carlo realization of the colour-magnitude diagram is shown in the left panel of Fig. 1. It is striking the high degree of similarity with the observational data, which are shown in the right panel of Fig. 1. Two clumps of stars are clearly visible in these diagrams. The bright one corresponds to unresolved binary stars, while the faint one corresponds to the pile-up of single white dwarfs owing to the combined effects of $^{22}$Ne sedimentation and carbon-oxygen phase separation.

Fig. 2 shows both the observed and the theoretical white dwarf luminosity functions. The solid line shows the average of $10^4$ Monte Carlo realizations corresponding to the age (8 Gyr), metallicity ([Fe/H]~+0.4) and distance modulus (13.44) of the cluster. Note the existence of two peaks in the white dwarf luminosity function, which are the direct consequence of the two previously discussed clumps in the colour-magnitude diagram. It is worth mentioning the very good agreement between the theoretical result and the observational data. Moreover, the main sequence turn-off and white dwarf ages are exactly the same, solving the age discrepancy of NGC 6791. Additionally, a $\chi^2$ analysis of the luminosity function reveals that, due to the narrowness of its two peaks, the cooling age determined in this way is very precise, being the age uncertainty $\pm 0.2$ Gyr.

To illustrate the importance of physical separation processes in Fig. 2 we also show, as a dotted line, the luminosity function obtained assuming that no physical separation processes occur and adopting the main-sequence turn-off age (8 Gyr). Clearly, the resulting luminosity function does not agree with the observational data. It could be argued that in this case the theoretical luminosity function could be reconciled with the observational data by simply decreasing the distance modulus by about 0.5 magnitudes. However, the same distance modulus should be then adopted to fit the main-sequence turn-off. If this were the case, we estimate that the main-sequence turn-off age would be ~12 Gyr, worsening the age discrepancy. Additionally, a distance modulus of 13.46±0.1 has been recently derived for NGC 6791 using eclipsing binaries[22], a totally independent and reliable method that does not make use of theoretical models. Thus, a large error in the distance modulus is quite implausible. Hence, the only possibility left to minimise the age discrepancy is to consider larger values of the metallicity, since isochrones with an enhanced metallicity have a fainter main sequence turn-off and, consequently, would result in a lower cluster turn-off age. However, to solve the age discrepancy a metallicity [Fe/H]~+0.7 would be needed. This metallicity is ~3σ from the most recent spectroscopic value[23]. Additionally, at this exceptionally high metallicity the predicted shape and star counts along the turn-off and sub-giant branch would be at odds with observations. Moreover, Fig. 2 shows that the fit to the observed luminosity function when the various physical separation processes are not included is very poor, since in addition to the mismatch in the locations of the peaks, the computed luminosity function cannot correctly reproduce the observed heights. The same occurs if only carbon-oxygen phase separation or only $^{22}$Ne sedimentation is included.

Based exclusively on the location of the cool end of the white dwarf sequence and not on the shape of the luminosity function we find that when both carbon-oxygen phase separation and $^{22}$Ne gravitational sedimentation are not taken into account, the age of the cluster turns out to be 6.0±0.2 Gyr. Thus, this type of cooling sequences, which are the most commonly used ones, can be safely discarded at the ~5σ confidence level, where σ≈0.4 Gyr is the uncertainty in the main sequence turn-off age. If only carbon-oxygen phase separation is considered the computed age of the cluster is 6.4±0.2 Gyr, so these sequences can also be excluded at the ~4σ confidence level, whereas if only

$^{22}$Ne sedimentation is taken into account we derive an age of 7.0±0.3 Gyr, which falls ~2.5σ off the main sequence turn-off age. Consequently, our results confirm unambiguously the occurrence of $^{22}$Ne sedimentation and strongly support carbon-oxygen phase separation in the deep interiors of white dwarfs. These findings have important consequences, as they prove the correctness of our understanding of the theory of dense plasmas and confirm that white dwarfs can be used as independent reliable chronometers.

**Acknowledgements** This research was partially supported by the MCINN, by the AGENCIA, by the Generalitat de Catalunya, by the STFC and by the CONICET. LGA also acknowledges a PIV grant from the AGAUR of the Generalitat de Catalunya. We are indebted to L. Bedin and co-authors for providing the observational colour-magnitude diagram of Fig. 1.

**Author information** Correspondence and requests for materials should be addressed to E.G. (garcia@fa.upc.edu).

**Figure legends**

**Figure 1 | Colour-magnitude diagrams of the white dwarfs in NGC 6791.** The left panel shows a typical Monte Carlo realization of the colour-magnitude diagram of NGC 6791. The blue dots are synthetic white dwarfs obtained using the procedure outlined in the main text and, thus, incorporating the photometric errors. A total of $\approx 850$ white dwarfs with magnitude smaller than $m_{F606W}=28.55^{mag}$ have been generated, the same number of white dwarfs observationally found[1,2,10]. The black dots show a theoretical white dwarf isochrone for 8 Gyr. Note the blue hook caused by the most massive white dwarfs of the cluster. The black lines are the observational selection area[10], white dwarfs outside this area are not considered. The right panel shows the observational white dwarf colour-magnitude diagram.

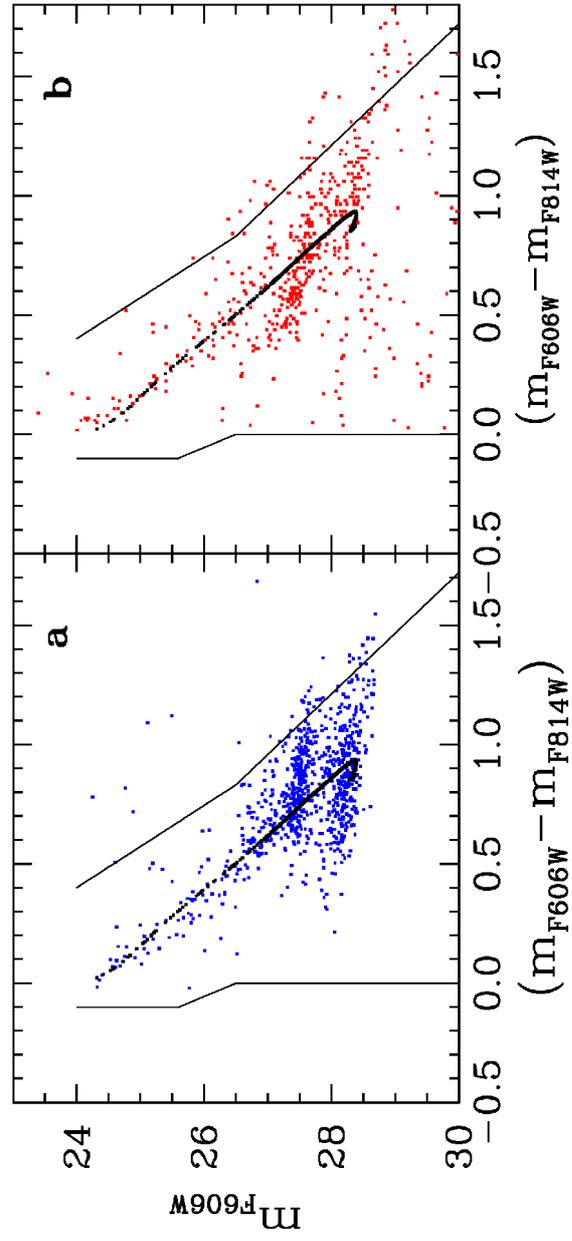

**Figure 2 | White dwarf luminosity function of NGC 6791.** The observational white dwarf luminosity function is shown as filled squares. The 1σ error bars correspond to the standard deviation[2]. The solid line is the average of $10^4$ Monte Carlo realizations corresponding to the age (8 Gyr), metallicity (0.04) and distance modulus (13.44) of NGC 6791. To illustrate the importance of physical separation processes, we also show the white dwarf luminosity function for the same age and assuming that no $^{22}$Ne sedimentation and no phase separation upon crystallisation occur (dotted line). The theoretical luminosity function is shifted to lower luminosities (larger magnitudes) to an extent that is incompatible with the observational data. The distance modulus required to fit the observations would be 13.0, a value considerably smaller than those observationally reported[1,2,22]. This distance modulus would imply a main-sequence turn-off age of 12 Gyr, worsening the age discrepancy[24]. Also shown at the top of the figure are the photometric error bars. Changes in the exponent of the initial mass function (of ±0.1) translate into small changes in the positions of the peaks (≤$0.02^{mag}$), well below the photometric errors ($0.15^{mag}$). As for the relationship between the mass of white dwarfs and the mass of their progenitors, the differences are also small (≤$0.04^{mag}$) when other recent relationships are adopted[25]. The same holds for reasonable choices of main-sequence lifetimes[26], in which case the differences are smaller than $0.02^{mag}$, or the duration of the burst of star formation (≤$0.04^{mag}$ when the duration of the burst is decreased to 0.1 Gyr).

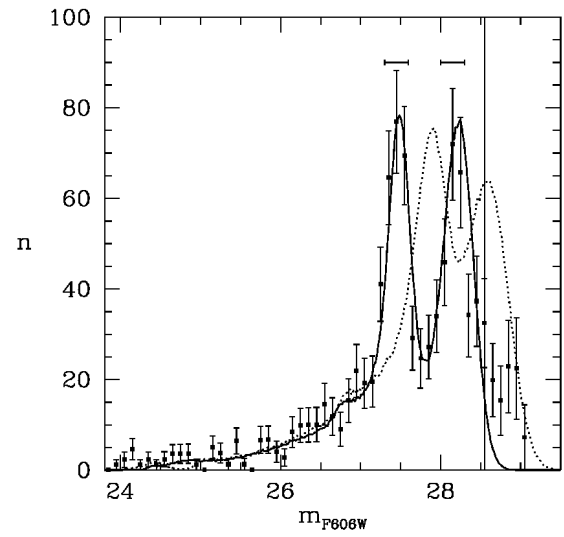